\documentstyle[prl,aps,amsmath,amssymb,epsfig]{revtex}

\newcommand{\etal}{\emph{et al}.}

\title{Neutron scattering study of
the field-induced soliton lattice in CuGeO$_3$}
\author{H. M. R\o{}nnow$^1$, M. Enderle$^2$, D. F. McMorrow$^1$
L.-P. Regnault$^3$,\\ G. Dhalenne$^4$, A. Revcolevschi$^4$,
A. Hoser$^5$, K. Prokes$^5$, P. Vorderwisch$^5$ and H. Schneider$^5$}
\address{$^1$Condensed Matter Physics and Chemistry Department,
Ris\o{} National Laboratory,
DK-4000 Roskilde, Denmark\\
$^2$Univ. Saarlandes, Saarbrucken, Germany,
$^3$CENG, Grenoble, France,
$^4$Laboratoire de Chimie des Solides, Universite de Paris Sud, Orsay, France.
$^5$BENSC, Berlin, Germany}

\date{\today}

\begin{document}
\twocolumn[\hsize\textwidth\columnwidth\hsize\csname
@twocolumnfalse\endcsname
\maketitle
\begin{abstract}
CuGeO$_3$ undergoes a transition from a spin-Peierls phase 
to an incommensurate 
phase at a critical field of $H_{\text c}\approx 12.5$ T.
In the high-field phase 
a lattice of solitons forms, 
with both  structural and  magnetic components, and these have been studied
using neutron scattering techniques.
Our results provide direct evidence for a long-ranged
magnetic soliton structure which has both  staggered and  uniform
magnetizations, and with amplitudes that are broadly in accord with
theoretical estimates. The  magnetic soliton
width, $\Gamma$, and the field dependence of the 
incommensurability, $\delta k_{\mathsc{sp}}$, 
are found to  agree well  with
theoretical predictions. 
\end{abstract}
\pacs{PACS numbers: }
]

A spin $\frac12$ antiferromagnetic (AF) chain in a deformable 3D lattice may,
at a characteristic temperature $T_{\mathsc{sp}}$, undergo a
spin--Peierls transition into a correlated dimerized 
groundstate characterized by a total spin $S_{\text{tot}}=0$.
The energy cost of deforming the
lattice is more than compensated for by the reduction in
magnetic energy, and a gap opens in the magnetic
excitation spectrum which reduces the influence of
zero--point fluctuations \cite{pytte74}.
This intriguing magneto--elastic phenomenon gives rise to several
interesting properties.
The groundstate is predicted to be a non-magnetic singlet separated
from an excited triplet ($S_{\text{tot}}=1$) 
by a gap of energy $\Delta_0=1.76\,T_{\mathsc{sp}}$.
The opening of this
gap produces anomalies in the temperature dependence of physical
quantities like the magnetic susceptibility, specific heat, etc.
Dimerization also results in a doubling of the unit cell along the
chain direction, which is reflected in diffraction experiments
by the appearance of satellite reflections at commensurate,
half-order positions.

Application of a magnetic field reduces the stability of the spin-Peierls
(SP) state against a state with finite magnetization
and, at a critical field $H_{\mathrm{c}}\simeq0.84\Delta_0/g\mu_B$, 
the system enters an incommensurate high-field 
phase\cite{cross79a,bray83}. 
Above $H_{\mathrm{c}}$, the development of a
uniform magnetization can be visualized as the breaking of a dimer bond,
thus allowing two spins to become parallel to the
magnetic field. 
The two $S^{z}=\frac{1}{2}$ states can be regarded as a
pair of domain walls with respect to the dimer order. 
The transverse part of the AF interaction delocalizes
the domain walls into solitons of finite width $\Gamma$, and
the longitudinal
part causes the two domain walls
to repel each other, thus forming an equally spaced lattice of solitons.
The resulting soliton has both  ferromagnetic and AF
components and, in addition, the lattice distortion also becomes
incommensurate. The average magnetization is proportional to the number of
solitons, and inversely proportional to the separation $L$.

Theoretically, the coupling of the spin chains to the lattice is
usually treated within the adiabatic approximation.
A continuum model has been developed by applying bosonization
techniques to the Jordan--Wigner transformed
Hamiltonian\cite{affleck90}.
This field-theoretical approach provides an analytic description
of the soliton structure in terms of the soliton spacing, 
$L/c=1/\delta k_{\mathsc{sp}}$, soliton width, $\Gamma$, and the amplitudes of 
the uniform, $m_{\mathrm{u}}$, and staggered,
$m_{\mathrm{s}}$, magnetizations\cite{nakano80,buzdin83,fujita84a}. 
The formalism has been extended to include next-nearest neighbour (nnn) and
inter-chain interactions\cite{zang97,dobry97}, but it is somewhat unclear
to what extent the results are affected by the various approximations
involved.
Monte Carlo\cite{feiguin97} and
density matrix renormalization group (DMRG)
calculations have also been performed\cite{meurdesoif99,uhrig99}. 
While the soliton shape 
matches the field theoretical solution, the DMRG
calculations show that the structural and magnetic soliton widths are
in general different to each other and field dependent.

As the high--field soliton lattice has both a structural and magnetic
component it is desirable to use a probe that is sensitive to both.
In this letter we report the results of an
investigation of the structure of the high--field phase of CuGeO$_3$ using
neutron scattering techniques.

Experimental investigations  of
the SP transition were initiated by the
work of Bray \etal\ on the organic compound
TTFCuS$_4$C$_4$(CF$_3$)$_4$\cite{bray83}.
Interest in SP systems intensified with the discovery by Hase
\etal\cite{hase93} that 
CuGeO$_3$ undergoes an SP transition at
$T_{\mathsc{sp}}=14$ K.
CuGeO$_3$ crystallises in the orthorhombic \emph{Pbmm} structure with
room-temperature lattice parameters of $a=4.80$ \AA, $b=8.47$ \AA\ and
$c=2.94$ \AA, and the $S=\frac12$ Cu$^{2+}$
moments form chains along the $c$-axis.
The field theory predictions depend on the spin gap $\Delta_0=2$ meV
and spin-wave velocity $v_s=16.6$ meV, 
which have been determined
using inelastic neutron scattering\cite{nishi94,regnault96,note_v_s}.
The inter-chain couplings $J_a=-0.08$ meV and $J_b=0.62$ meV
indicate that the system is in fact only quasi-1D\cite{cowley96}.
Along the chain, both an alternating exchange coupling
$J^\pm_c=(1\pm0.014)\times14$ meV and a significant nnn coupling
$J_c'\simeq0.35J_c$ are needed to model the susceptibility, saturation
field and excitation spectrum\cite{riera95}.

In zero-field the
expected doubling of the unit cell along the $c$ direction
below $T_{\mathsc{sp}}$
has been confirmed by
electron\cite{kamimura94}, X-ray\cite{pouget94} and
neutron\cite{hirota94,braden96} 
diffraction experiments.
The strongest satellite reflections appear at
$(\frac{h}2,k,\frac{l}2)$ with $h$, $k$ and $l$ odd, which indicates that
along $a$ and $b$ adjacent chains dimerize in anti-phase.
Hirota \etal\cite{hirota94} and Braden \emph{et al.}\cite{braden96}
have solved the detailed structure of the  SP phase,
which is characterized by 
the propagation wavevector $k_{\mathsc{sp}}=(\frac12,0,\frac12)$ and
three displacement parameters:
$u_c^{\text{Cu}}/c=0.00192$, $u_a^{\text{O2}}/a=0.00198$,
$u_b^{\text{O2}}/b=0.00077$.

Several experiments have been reported on the high--field phase,
including X-ray scattering \cite{kiryukhin96b}, 
thermal expansion\cite{lorenz98}, FIR spectroscopy \cite{loosdrecht96b},
NMR\cite{horvatic99}, ESR and magnetization \cite{brill94,hori95}.
The predicted incommensurability of the structure
was observed in an X-ray experiment
\cite{kiryukhin96b} as a splitting of the satellite
reflection into $(\frac72,1,\frac52\pm\delta k_{\mathsc{sp}})$, and
the soliton width was estimated to be $\Gamma=(13.6\pm0.3)c$.
Using copper NMR, Horvati\'{c} \etal{} \cite{horvatic99} have 
determined the staggered and uniform magnetizations to 
be $m_{\mathrm s}=0.026$ and 
$m_{\mathrm u}=0.023$, respectively at 14.5 T.
(By definition a spin of $\frac12$ corresponds to $m=\frac12$.)
Magnetization measurements give a value of 0.018 for $m_{\mathrm u}$.
The theoretical values are $m_{\mathrm s}=0.11 \leftrightarrow$ 0.14 and
$m_{\mathrm u}$=0.023 \cite{nakano80,buzdin83,fujita84a,zang97,dobry97,uhrig99}.
Thus although there is good agreement between  theory
and experiment for $m_{\mathrm u}$, the staggered moment observed in
NMR is significantly lower  than the theoretical value, a point we
will return to later.

Although neutron scattering is an obvious way to attempt to study the magnetic
soliton lattice, it has until recently not been possible to apply
this technique above 12 T.
In CuSi$_{.03}$Ge$_{.97}$O$_3$, where $H_{\mathrm{c}}$ is reduced to 11.7 T,
Grenier \etal\cite{grenier98} have measured the intensity of five
satellite reflections upon entering the high--field phase.
No clear evidence of any magnetic scattering was found.
The soliton width at 12 T was deduced to be 13.5$c$, in agreement
with the X-ray data on pure CuGeO$_3$.

Here, we report on a neutron scattering study of the
high--field phase in pure CuGeO$_3$, performed at HMI, Berlin, 
using a 14.5 T vertical
field cryomagnet with a base temperature of 1.6 K.
The single crystal of dimensions $3\times7\times16$ mm$^3$
was grown by the floating zone method. In total
three different instruments were used, each chosen to reveal a
particular aspect of the phase transition.
Due to the slightly anisotropic $g$-tensor\cite{pilawa97}, the critical
field depends on the orientation of the applied magnetic
field. 

The field dependence of the incommensurability  was investigated using the
thermal triple--axis spectrometer E1 with a collimation of
20'-40'-20'-80', and a wavelength of $\lambda=2.425$ \AA.
By performing scans along $c^\ast$ through five satellite reflections
$(\frac12,1,\frac12)$, $(\frac12,3,\frac12)$, $(\frac32,3,\frac12)$,
$(\frac12,1,\frac32)$ and $(\frac12,3,\frac32)$, we observed the
expected splitting.
\begin{figure}
\includegraphics[scale=0.4]{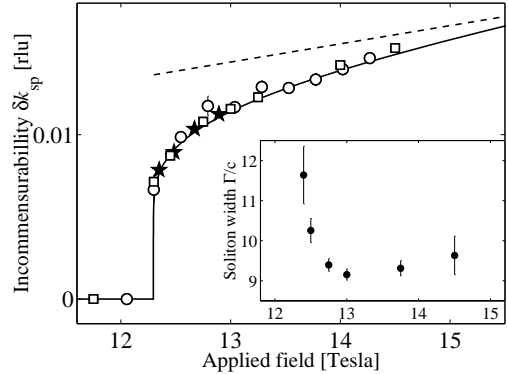}
\caption{The measured splitting $\delta k_{\mathsc{sp}}$ of
  $(\frac12,1,\frac12)$ ($\Box$) and $(\frac12,3,\frac12)$ ($\circ$),
  compared to X-ray data\protect\cite{kiryukhin96b}
  ($\bigstar$), and to the theories of Cross and Fisher\protect\cite{cross79a}
  (dashed line), and Buzdin \etal\protect\cite{buzdin83}
  (solid line with
  $\Delta_0/v_s=0.13$). For clarity, only a representative
  selection of the X-ray data is shown, and all of the data has been
  scaled to the $g$ factor corresponding to the $(h,2h,l)$ orientation.
The insert shows the soliton width $\Gamma$ as obtained from the
relative intensity of the 3rd harmonics.}
\label{fig:splitting}
\end{figure}
\noindent
As shown in Fig.\ \ref{fig:splitting}, the field dependence of
$\delta k_{\mathsc{sp}}$ is consistent with the X-ray results in the
region of overlap, but our data extend to much higher
fields. 
Field theory \cite{buzdin83} predicts that the incommensurability of
the soliton lattice is given by
$1/\delta k_{\mathsc{sp}}=\frac{2v_s}{\Delta_0}\ln\frac{8H_{\mathrm{c}}}{H-H_{\mathrm{c}}}$.
Using the upper limit value of 0.13 for the experimentally determined
ratio $\Delta_0/v_s=0.12\pm0.01$, we obtain
perfect agreement with our data.
At the
highest fields, the linear field dependence $\delta
k_{\mathsc{sp}}=g\mu_BH/(2\pi v_s)$ is approached\cite{cross79a}.

One striking feature of this data set is that there 
was a tremendous increase in intensity of the low $Q$ 
satellites. This appears difficult to explain on
the basis of a purely structural
distortion, for which the intensity increases in proportion to $Q^2$.
One possible explanation could be that the magnetic soliton lattice has
the same propagation vector as the lattice soliton.
At first sight this seems unlikely, as the minimum energy
gap in the spin-wave spectrum is located at $(0,1,\frac12)$, which is
also the propagation vector of the AF structure found
in doped systems\cite{lussier95}.
It should be noted, however, that the magnetic coupling in the $a$ 
direction is weak, of order 1~K, and could be overcome by stronger 
lattice forces.
To exclude the possibility that there is magnetic scattering from the
staggered component at wavevectors other than
$(\frac{h}2,k,\frac{l}2)$, an exhaustive search for additional
satellite reflections that may be
associated with long range magnetic order was conducted.
From this an upper limit can be put on any Fourier component at
$(0,1,\frac12)$, $(0,3,\frac12)$ and $(0,0,\frac12)$ of $0.005\mu_B$
at 14.5 T.

To investigate whether in fact there is a magnetic contribution to the
low-$Q$ satellites, a second
experiment was undertaken. 
The  two-axis spectrometer E4 was used with a collimation of
40'-40'-40' and $\lambda=1.2205$ \AA, which made 
14 satellite reflections  accessible in the
$(h,2h,l)$-plane (see Table \ref{tab:f2}). Rocking curves
of each reflection were collected at
2~K in zero field and 14.5~T.
An additional data set was taken at 20~K 
(above $T_{\mathsc{sp}}=14.1$~K) and 0~T to
determine the $\lambda/2$ contamination.
The widths of all measured peaks were in complete
agreement with an analytic resolution calculation, and
this was used to correct  the integrated intensities of the  measured
satellite peaks.
The overall scale factor that converts the satellite intensities to
structure factors can be found by accurately determining the
structure factors of the main Bragg peaks. However, this process
is often problematic, as it involves additional corrections
for extinction, absorption, and in our case
$\lambda/2$ contamination. To avoid these problems,
we chose instead an overall scale factor such that the sum
of the displacement parameters obtained when fitting our zero--field
data matched the sum of the parameters obtained
in the detailed study by Braden \etal\cite{braden96}.
The relative size of our zero--field displacement parameters
$u_z^{\mathrm{Cu}}/c=0.00204\pm0.00014$,
$u_x^{\mathrm{O2}}/a=0.00182\pm0.00010$ and
$u_y^{\mathrm{O2}}/b=0.00079\pm0.00009$ 
agree well with the values reported by Braden \etal.

\begin{figure}
\includegraphics[scale=0.4]{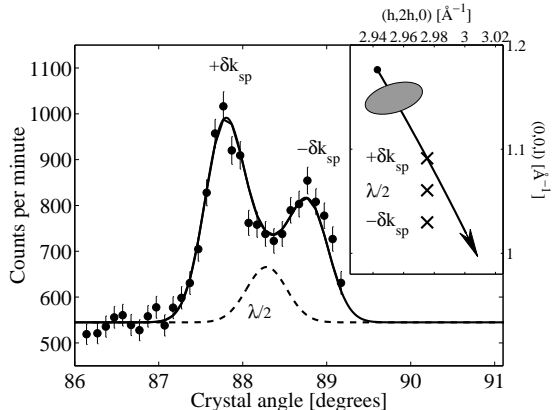}
\caption{Rocking curve through $(\frac32,3,\frac12+\delta k_{\mathsc{sp}})$ at 14.5 T,
  2 K. The finite instrument resolution picks up both satellites
  $(\frac32,3,\frac12\pm\delta k_{\mathsc{sp}})$, and the $\lambda/2$ background at
  $(\frac32,3,\frac12)$.
  The solid line shows a fit of three resolution convoluted peaks.
  The insert displays the reciprocal space trajectory of the
  scan and the resolution ellipse. 
  The position of each peak is marked with a cross.}
\label{fig:e4data}
\end{figure}
\noindent

The same scale factor was used for the data taken in the
high--field phase at 14.5 T (Table \ref{tab:f2}).
The resolution obtainable with the short wavelength
was insufficient to completely separate the two
satellites $(\frac{h}2,h,\frac{l}2\pm\delta k_{\mathsc{sp}})$ and the
$\lambda/2$ background at all $(\frac{h}2,h,\frac{l}2)$.
This  is illustrated
in the scan through $(\frac32,3,\frac12+\delta k_{\mathsc{sp}})$ shown in Fig.\
\ref{fig:e4data}.
The data were fitted to three peaks convoluted with the calculated
resolution. The $\lambda/2$ contribution was fixed to the value
determined at 20 K and 0 T, and the two satellites were given equal
amplitude.
With the satellite amplitude as the only fitting parameter, all fourteen
reflections gave excellent fits as shown  by the solid
line in Fig.\ \ref{fig:e4data}.
The resulting structure factors are listed in  Table \ref{tab:f2}.

\begin{table}
\begin{tabular}{ccc|rrrr}
&&& \multicolumn{2}{c}{O T} & \multicolumn{2}{c}{14.5 T}\\
$2h$ & $k$ & $2l$ &
$|F|^2_{\mathrm{exp}}$ &
$|F|^2_{\mathrm{fit}}$ &
$|F|^2_{\mathrm{exp}}$ &
$|F|^2_{\mathrm{fit}}$ \\\hline
1 & 1 & 1 &  4 $\pm$ 1 & 1 & 40 $\pm$ 1 & 41\\
3 & 3 & 1 &  3 $\pm$ 2 & 1 & 31 $\pm$ 1 & 29\\
1 & 1 & 3 &  26 $\pm$ 5 & 16 & 33 $\pm$ 1 & 30\\
3 & 3 & 3 &  36 $\pm$ 3 & 34 & 29 $\pm$ 3 & 26\\
5 & 5 & 1 &  30 $\pm$ 4 & 20 & 16 $\pm$ 1 & 21\\
1 & 1 & 5 &  35 $\pm$ 11 & 34 & 14 $\pm$ 5 & 20\\
5 & 5 & 3 &  83 $\pm$ 4 & 89 & 30 $\pm$ 2 & 28\\
3 & 3 & 5 &  18 $\pm$ 6 & 16 & 12 $\pm$ 2 & 15\\
7 & 7 & 1 &  86 $\pm$ 8 & 90 & 25 $\pm$ 5 & 37\\
5 & 5 & 5 &  13 $\pm$ 8 & 0 & 10 $\pm$ 1 & 9\\
1 & 1 & 7 &  89 $\pm$ 10 & 81 & 16 $\pm$ 2 & 21\\
7 & 7 & 3 &  218 $\pm$ 14 & 208 & 87 $\pm$ 7 & 59\\
3 & 3 & 7 &  85 $\pm$ 17 & 116 & 28 $\pm$ 3 & 25\\
7 & 7 & 5 &  25 $\pm$ 32 & 21 & 0 $\pm$ 7 & 17\\
\multicolumn{3}{c}{$\chi^2$} &
\multicolumn{2}{c}{2.8} &
\multicolumn{2}{c}{6.5}
\end{tabular}
\caption{Measured and fitted structure factors in
  units of $10^{-27}$ cm$^{-2}$.}
\label{tab:f2}
\end{table}

As anticipated, it was not possible to obtain satisfactory  fits of
the high-field structure factors
by assuming a structural distortion alone. Instead, a magnetic
Fourier component of variable size and direction 
with the magnetic form-factor of Cu$^{2+}$ was included in the fit.
The resulting displacement parameters
$u_z^{\mathrm{Cu}}/c=0.00158\pm0.00026$,
$u_x^{\mathrm{O2}}/a=0.00196\pm0.00036$ and
$u_y^{\mathrm{O2}}/b=0.00061\pm0.00025$ are not significantly  different
from the zero field parameters.
The direction of the magnetic moments was found to be
parallel/anti-parallel to the field, in agreement with 
theoretical expectations.
The Fourier component was determined to be 
$\mu(q=\delta k_{\mathsc{sp}})=0.098\pm0.003\mu_B$ at 14.5 T.

\begin{figure}
\includegraphics[scale=0.4]{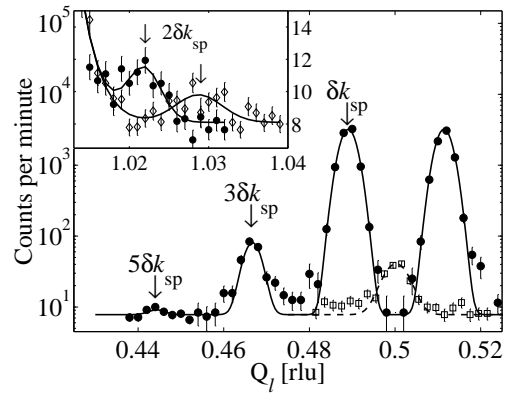}
\caption{Scans along [00$l$]
through $(\frac12,1,\frac12)$ at 1.6 K and 0 T (squares),
and 13 T (filled circles).
The resolution limited peaks
were fitted to  Gaussian line
shapes.
The higher harmonics $n\delta k_{\mathsc{sp}}$ are marked by arrows.
The insert shows scans through $(0,0,1+2\delta k_{\mathsc{sp}})$ 
at 13 T (filled circles) and
14.5 T (diamonds).}
\label{fig:typscan}
\end{figure}

Having established that the $(\frac12,1,\frac12\pm\delta
k_{\mathsc{sp}})$ 
reflection is almost
completely magnetic in origin, we performed a high-resolution
experiment in order to investigate the detailed structure of the
magnetic soliton lattice.
Using the cold triple axis spectrometer V2 with $\lambda=5.236$~\AA\
and 20' collimation gave a resolution of 0.009 \AA$^{-1}$.
Fig.\ \ref{fig:typscan} shows scans through $(\frac12,1,\frac12)$
along $[00l]$ at 0
T and at 13 T. Along with the dramatic increase in intensity, the 13 T
scan reveals both third and fifth harmonics, which reflect the
non-sinusoidal shape of the magnetic solitons.

From field theory, an analytical solution 
for the magnetization 
$\langle S^z(l)\rangle
=m_{\mathrm{s}}(-1)^l\text{cn}(lc/\Gamma k,k)+m_{\mathrm{u}}\text{dn}(lc/\Gamma k,k)$ 
at site $l$
with amplitudes
$m_{\mathrm{s}}=\sqrt{\frac{\Delta_0}{2\pi v_s}}$ and 
$m_{\mathrm{u}}=\frac{\Delta_0}{2\pi kv_s}$
is given in terms of the Jacobian elliptic functions cn and dn of
modulus $k$, where $L/\Gamma=4kK(k)$ and $K(k)$ is the complete
elliptic integral of first kind\cite{nakano80,buzdin83,fujita84a,zang97}.
The $n$th Fourier components of the staggered and uniform parts of the
magnetic structure are given by
$\frac{\pi}{kK(k)}\frac{\kappa^{n/2}}{1+\kappa^{n}}$
($n$ odd) and 
$\frac{\pi}{K(k)}\frac{\kappa^{n/2}}{1+\kappa^{n}}$
($n>0$ even) respectively, where
$\kappa=e^{-\pi K(\sqrt{1-k^2})/K(k)}$.
From the ratio $\kappa^2/(1-\kappa+\kappa^2)^2$ between the 
intensities of the third and the first
harmonic, we determine the modulus $k$ and the soliton width
$\Gamma/c=1/(4\delta k_{\mathsc{sp}}kK(k))$, which is shown in the insert of
Fig.\ \ref{fig:splitting}.
Above the transition, $\Gamma/c$ quickly decreases from $11.5$ to a
value of $9$, followed by a slight increase with field,
a feature which is also apparent in
DMRG calculations\cite{uhrig99}.
Except for the rapid change just above $H_c$, our data
are in fair agreement with the
constant value $\Gamma/c=v_s/\Delta=8.3$
predicted by field theory, but
differs from the values $\Gamma/c\sim13.5$ 
obtained from X-ray and neutron
scattering on the Si doped system.
We believe that
this is because the latter experiments measure the width of the
structural soliton, which in the presence of nnn coupling is larger
than the magnetic soliton width\cite{uhrig99}.
Re-analysing the X-ray data, by allowing for a field dependent soliton
width, leads to a similar decrease $\Gamma/c=15.1\rightarrow12.9$ 
with increasing
field.

Knowing the precise soliton shape, the amplitude 
of the staggered moment
can be determined
from the Fourier component:
$m_{\mathrm{s}}=\frac{kK(k)}{\pi}\frac{1+\kappa}{\sqrt{\kappa}}
\frac{\mu(q=\delta k_{\mathsc{sp}})}{g}
=0.097\pm0.003$, where $g=2.19$ for $B\perp(h,2h,l)$\cite{pilawa97}.
This value is in reasonable agreement with theoretical
estimates cited earlier, but is considerably larger than the value derived
from NMR.
If, as has been suggested, the NMR value is reduced by quantum
fluctuations in the groundstate\cite{horvatic99,uhrig99}, then these
would also reduce the value of $m_s$ observed in our neutron
scattering experiments.
The NMR and neutron scattering data can be reconciled if there are
fluctuations fast compared to the time scale probed by NMR ($\sim10^{-4}$
THz) but slow compared to that probed by the neutron scattering
($\sim$THz).

The uniform part of the magnetic structure should give rise to even
harmonics around integer positions in reciprocal space.
Indeed, second-order harmonics were found near the
(0,0,1) Bragg reflection, as shown in the insert of Fig.\
\ref{fig:typscan}.
To extract the absolute structure factor, we scale the integrated
intensity to the $(\frac12,1,\frac12\pm\delta k_{\mathsc{sp}})$ reflection
measured in the same configuration, correcting for the different
resolution volumes.
The resulting Fourier component of the magnetic structure is 
$\mu(0,0,1+2\delta k_{\mathsc{sp}})=0.0046 \mu_B$ at 14.5 T.
The amplitude of the uniform component is then given by
$m_{\mathrm{u}}=\frac{K(k)}{\pi}\frac{1+\kappa^2}{\kappa}
\frac{\mu(q=2\delta k_{\mathsc{sp}})}{g}=0.019\pm0.003$,  about 25\% lower
than the field theory and  NMR results.
It should be noted that
although the individual values of $m_{\mathrm s}$ and $m_{\mathrm u}$
are sensitive to the details of any normalisation procedure,
the ratio $m_{\mathrm s}/m_{\mathrm u}=5.1\pm 0.8$ is not, 
and is seen to be in 
good agreement with  the theoretical ratio $m_{\mathrm s}/m_{\mathrm u}=5.4.$. 

In summary, neutron scattering has been used to determine the
nature of the soliton lattice in the
high-field phase of the spin-Peierls compound CuGeO$_3$.
The results provide a detailed description of the soliton lattice in
terms of incommensurability, $\delta k_{\mathsc{sp}}$, soliton width, $\Gamma$,
and the amplitudes $m_{\mathrm{s}}$  and $m_{\mathrm{u}}$ of both the
staggered and uniform parts of the magnetization. 

We wish to thank R. A. Cowley, B. Keimer, B. Lebech, K. Lefmann, B. Lake
and V. J. Emery for valuable discussions, 
M. Meissner and P. Smeibidl for handling the cryomagnet. 
The experiments were performed at the HMI, Berlin with support from EC
through TMR--LSF contract ERBFMGECT950060. We acknowledge support for
HMR from the Danish Research Academy and for ME from BMBF.


\begin{thebibliography}{10}
\bibitem{note_v_s}
The spin-wave velocity $v_s$ is defined such that the dispersion follows
  $\hbar\omega(q)=\sqrt{\Delta_0+(v_sq)^2}$, and it value is uniquely detemined
  from neutron scattering data\cite{regnault96}.
\end{thebibliography}

\end{document}